\documentclass[prb,aps,twocolumn,superscriptaddress,showpacs,floatfix]{revtex4-2}

\usepackage{float}
\usepackage[pdftex]{graphicx}
\usepackage{epstopdf}
\usepackage[colorlinks=true,linkcolor=black, citecolor=blue, urlcolor=blue, 
    unicode=true]{hyperref}
\usepackage{amsmath}
\usepackage{subfigure}
\usepackage{fixltx2e}
\usepackage{dcolumn}
\renewcommand\labelenumi{(\roman{enumi})}
\renewcommand\theenumi\labelenumi
\newcommand{\sr}{Sr$_3$Ir$_2$O$_7$F$_2$}
\newcommand{\jeff}{$J$\textsubscript{eff}}

\begin{document}

\title{Spin-orbit excitons and electronic configuration of the $5d^4$ insulator Sr$_3$Ir$_2$O$_7$F$_2$}

\author{Zach Porter}
\affiliation{Materials Department, University of California, Santa Barbara, California 93106-5050, USA}

\author{Paul M. Sarte}
\affiliation{Materials Department, University of California, Santa Barbara, California 93106-5050, USA}

\author{Thorben Petersen}
\affiliation{Institute for Theoretical Solid State Physics, Leibniz-IFW Dresden, Helmholtzstra{\ss}e 20, D-01069 Dresden, Germany}

\author{Mary H. Upton}
\affiliation{Advanced Photon Source, Argonne National Laboratory, Argonne, Illinois 60439, USA}

\author{Liviu Hozoi}
\affiliation{Institute for Theoretical Solid State Physics, Leibniz-IFW Dresden, Helmholtzstra{\ss}e 20, D-01069 Dresden, Germany}

\author{Stephen D. Wilson}
\affiliation{Materials Department, University of California, Santa Barbara, California 93106-5050, USA}
\email[email: ]{stephendwilson@ucsb.edu}

\date{\today}

\begin{abstract}
Here we report on the low-energy excitations within the paramagnetic spin-orbit insulator Sr$_3$Ir$_2$O$_7$F$_2$ studied via resonant inelastic X-ray scattering, \textit{ab initio} quantum chemical calculations, and model-Hamiltonian simulations. This material is a unique $d^{4}$ Ir$^{5+}$ analog of Sr$_3$Ir$_2$O$_7$ that forms when F ions are intercalated within the SrO layers spacing the square lattice IrO$_{6}$ bilayers of Sr$_3$Ir$_2$O$_7$. Due to the large distortions about the Ir$^{5+}$ ions, our computations demonstrate that a large single-ion anisotropy yields an $S$=1 ($L{\approx}$1, $J{\approx}$0) ground state wave function.  Weakly coupled, excitonic modes out of the $S_z$=0 ground state are observed and are well-described by a phenomenological spin-orbit exciton model previously developed for $3d$ and $4d$ transition metal ions.  The implications of our results regarding the interpretation of previous studies of hole-doped iridates close to $d^{4}$ fillings are discussed. 
\end{abstract}
\maketitle

%%%%%%%%%%%%%%%%%%%%%%%%%%%%%%%%%%%%%%%%%%%%%%%%%%%%%

A considerable amount of theoretical~\cite{Khaliullin2013, Meetei2015, Kunes2015, Bhowal2015, Chaloupka2016, Svoboda2017, Kim2017, Geffroy2019} and experimental~\footnote{For an experimental review see Ref. \citenum{Cao2018}. Some more recent examples are Refs. \citenum{Kusch2018, Davies2019, Takahashi2021}.} work has been dedicated to exploring the magnetism of the \jeff=0~ground state and its potential for excitonic magnetism. The driving idea is that strong spin-orbit coupling (SOC) native to heavy transition metal ions in high-symmetry crystal field environments can lead to \jeff=0~ground states, and the Van Vleck-type spin-orbit excitons that result above this ground state (e.g. $J$=0 to $J$=1) can condense into so-called `excitonic' magnetic order \cite{Khaliullin2013}. This can engender a quantum critical point with high energy scale fluctuations \cite{Meetei2015}. And, for some material realizations, the condensed phase itself may support novel spin-liquid states.

It is rare to find a materials platform that enables study of spin-orbit excitonic states upon changing the ground state wave function from magnetic \jeff=1/2~to nonmagnetic \jeff=0, all while preserving the lattice topology and metal site ion.  Excitingly, recent studies have shown that previously established topochemical transformation techniques and intercalation pathways developed for Ruddelsden-Popper phases in $3d$ and $4d$ transition metal oxides \cite{zhang2016directed, PhysRevB.62.3811} can also be leveraged on their $5d$-electron cousins \cite{Peterson2018, PhysRevMaterials.4.013403}.  This provides the opportunity to control the oxidation state in $5d^5$ iridates with \jeff=1/2~ground states and drive them into $5d^4$ \jeff=0~states, while preserving the metal ion character and lattice topology. 

Of particular interest is Sr$_3$Ir$_2$O$_7$F$_x$ with $0{\leq}x{\leq}2$, a $d^5$ iridate which can be driven into the $d^4$ state via fluorine intercalation \cite{Peterson2018}. This fluorine system presents several advantages for studying $5d^4$ physics over other modifications to Sr$_3$Ir$_2$O$_7$. Previous studies of Sr$_3$Ir$_2$O$_7$ when alloyed with $4d^4$ Ru have resolved an anomalous hardening of magnon modes combined with a narrowing of the excitation bandwidth as the Sr$_3$Ru$_2$O$_7$ endpoint is approached \cite{Schmehr2019}, similar to results for the monolayer analog Sr$_2$(Ir,Ru)O$_4$ \cite{Cao2017}. Resolving a crossover into a pure $d^4$ excitonic state was prohibited by the diminishing fraction of Ir needed for conventional hard resonant X-ray measurements as well as by the intrinsic disorder broadening native to the Ir/Ru alloying in the system. Single crystals of Sr$_3$Ir$_2$O$_7$F$_2$ potentially provide a unique new window into this problem by circumventing both of these roadblocks. 

Here we present a study of spin-orbit exciton dynamics in the $5d^4$ system Sr$_3$Ir$_2$O$_7$F$_2$.  By using RIXS measurements combined with \textit{ab initio} quantum chemistry calculations and a phenomenological spin-orbit exciton model, we demonstrate that Sr$_3$Ir$_2$O$_7$F$_2$ realizes a spin-orbit singlet $S{=}1$ ($L{\approx}1$, $J{\approx}0$) ground state whose excitation spectrum is governed by the interplay between strong SOC and strong distortions within the local ligand fields about the Ir-sites.  Weakly coupled spin-orbit excitons and $d{-}d$ excitations are observed and establish the energy scales of SOC, coupling strengths, and crystal field splitting.  The implications for interpreting the anomalous excitation spectrum observed in heavily substituted Sr$_3$(Ir$_{1-x}$Ru$_x$)$_2$O$_7$ alloys are discussed. 

%%%%%%%%%%%%%%%%%%%%%%%%%%%%%%%%%%%%%%%%%%%%%%%%%%%%%

\begin{figure}
\includegraphics[trim=5mm 7mm 59mm 2mm, clip,width=0.42\textwidth]{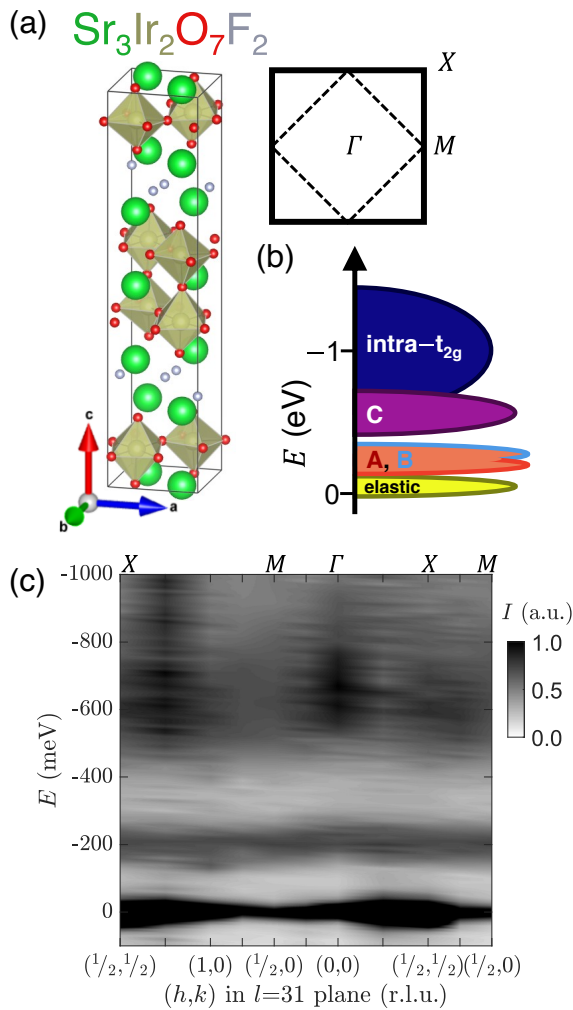}
\caption{(a) Orthorhombic $Bbcb$ unit cell (left) and quasi-2D Brillouin zone (right) for \sr. Ir$-$O bond lengths are appreciably compressed along $c$ near the fluorine planes. (b) Illustration of the excitations observed in RIXS measurements. Details are in the text. (c) RIXS false-color map of raw intensities in the quasi-2D Brillouin zone. Ticks indicate the $\mathbf{Q}$ positions where spectra were measured, and the map was generated by interpolation.} \label{fig:fig1}
\end{figure}

Sr$_3$Ir$_2$O$_7$F$_2$ crystals were prepared using the methods described elsewhere \cite{Peterson2018}. The exact synthetic conditions and further characterization are described in the supplemental information \cite{ESI}.  Resonant inelastic X-ray scattering (RIXS) measurements were performed at 8 K on Beamline 27-ID-B of the Advanced Photon Source at Argonne National Laboratory. The incident photons were tuned to the Ir $L_3$ absorption edge ($E{=}11.215$\:keV) and final energies were selected with the Si (4,4,8) reflection of a spherical analyzer crystal array in a horizontal scattering geometry \cite{Shvydko2013}.  Excitations were mapped primarily in the quasi-2D $l$=31 Brillouin zone (BZ) with an energy resolution $\approx$35 meV.  This $l$ value was chosen because it is far from Bragg peaks and near $2\theta$=90$^\circ$ where Thomson scattering (i.e. elastic charge scattering) is minimized.  Momentum space positions are indexed using an orthorhombic $Bbcb$ unit cell with lattice parameters $a$=5.45\:{\AA}, $b$=5.51\:{\AA}, and $c$=24.21\:{\AA}; see Figure \ref{fig:fig1}. This simplification from the proper $C2/c$ cell was chosen for comparison to other quasi-2D perovskite variants. Due to twin structural domains, we do not distinguish between $a$ and $b$ axes. 

\textit{Ab initio} quantum chemical calculations were performed on both single-octahedron [IrO$_{6}$]$^{7-}$ and two-octahedra [Ir$_2$O$_{11}$]$^{12-}$ clusters \cite{ESI}, embedded in point charge fields (PCFs) created by using experimentally determined lattice positions \cite{Peterson2018} as input for the \textsc{Ewald} program \cite{Derenzo2000, Klintenberg2000}.  Excited state energies were calculated on the basis of \textit{ab initio} complete active space self-consistent field (CASSCF) theory \cite{Roos1987, Helgaker2012}, considering the Ir $t_\text{2g}$ atomic orbitals in active spaces of the sizes CAS(4e,3o) and CAS(8e,6o) for single-octahedron and two-octahedra clusters, respectively.  Subsequently, dynamical correlation was treated in the scheme of multi-reference configuration interaction (MRCI), with the Ir $4f$, Ir $5s$, Ir $5p$, and O $2p$ orbitals correlated in the CI part.  All calculations were performed with the program package ORCA v5.0 \cite{Neese2022} and further details are in the supplemental information \cite{ESI}.

Looking first at Figure \ref{fig:fig1}, the structure of Sr$_3$Ir$_2$O$_7$F$_2$ is shown in panel (a).  IrO$_6$ octahedra are asymmetrically compressed, with the O ions closest to intercalated F ions pushed inward toward the Ir ions.  Figure \ref{fig:fig1}(b) provides a qualitative illustration of the spectral features and relative energy scales probed by RIXS measurements at the Ir $L_3$-edge in Sr$_3$Ir$_2$O$_7$F$_2$, and Figure \ref{fig:fig1}(c) shows a momentum-energy transfer map of RIXS intensities measured across the quasi-2D Brillouin zone.

\begin{figure}
\includegraphics[trim=4mm 9mm 75mm 3mm, clip,width=0.42\textwidth]{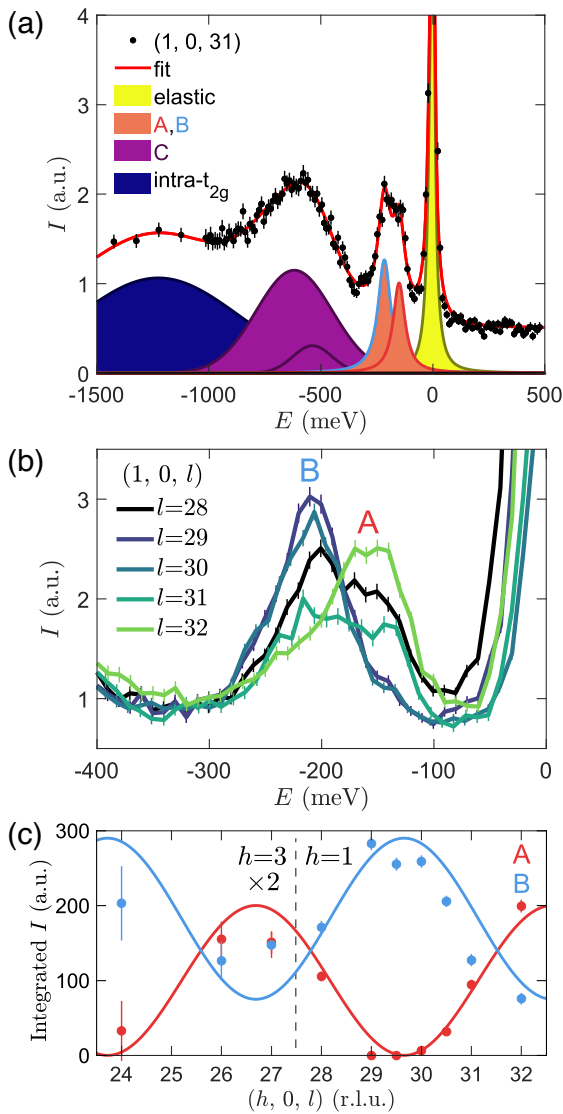}
\caption{(a) Representative RIXS spectrum (black dots) collected at \textbf{Q}=(1, 0, 31). Fits to the spectra (solid red lines) utilize the spectral components discussed in the text in addition to a constant background term. (b) Raw RIXS spectra at (1, 0, $l$) show the change in spectral weight of the A and B features (red and blue, respectively). (c) Integrated intensities of the Lorentzian fits to these features reveal sinusoidal dependence in $l$. The functional form of the red line is $\sin^2(\pi l d/c)$, as is described in the text. Note that for $l{<}28$ the fitted scans were measured at (3, 0, $l$) which was closer to normal incidence, where self-absorption effects were larger due primarily to the scattering geometry. The $h$=3 scans were scaled by a factor of 2 to make the intensity values comparable.}
\label{fig:fig2}
\end{figure}

All modes observed are weakly dispersive, and we label the features as elastic at energy loss $E$=0\:meV, A at $E{\approx}$170\:meV, B at $E{\approx}$220\:meV, and C at $E{\approx}$500-900\:meV. A, B, and C are spin-orbit excitons. A representative fit parameterizing these features is shown in Figure \ref{fig:fig2}(a), where the elastic line was fit to a Voigt function, peaks A and B were fit to Lorentzians, and other features were empirically fit to Gaussians. The key features parameterized within this spectrum are the A and B peaks, which we will demonstrate are spin-orbit excitons out of the $S{=}1, \; J{=}0$ ground state. 

An important first step is to demonstrate the presence of two distinct exciton modes near 200 meV, which are nominally split by the large noncubic distortion of the IrO$_6$ octahedra. Figure \ref{fig:fig2}(b) shows the RIXS spectra about this energy region at the 2D zone center with varying $l$-values.  Changing the value of $l$ modulates the intensities of the A and B modes; mode A can be mostly isolated at $l{=}32$ and mode B at $l{=}29$.  The integrated intensities of the A and B modes are plotted as a function of $l$ in Figure \ref{fig:fig2}(c), where the intensity of the A mode is well-described by the form $\sin^2(\pi l d/c)$, with $d$ the bilayer Ir-Ir spacing and $c$ the lattice constant. For the B mode, the modulation of the intensity is $\pi/2$ out of phase with an added constant background. We attribute this sinusoidal behavior of A and B to a double-slit-like interference for two different excited states that are delocalized across the bilayer. This interference effect results from the emission process in RIXS, \cite{Ma1995}, and was invoked to explain the scattering for Sr$_3$Ir$_2$O$_7$~\cite{mazzone2022antiferromagnetic}
and for dimer compounds like Ba$_3$CeIr$_2$O$_9$~\cite{Revelli2019, Wang2019}. A more detailed description of the physical origin and alternative explanations are provided in the supplementary material \cite{ESI}.

\begin{figure}
\includegraphics[scale=0.32]{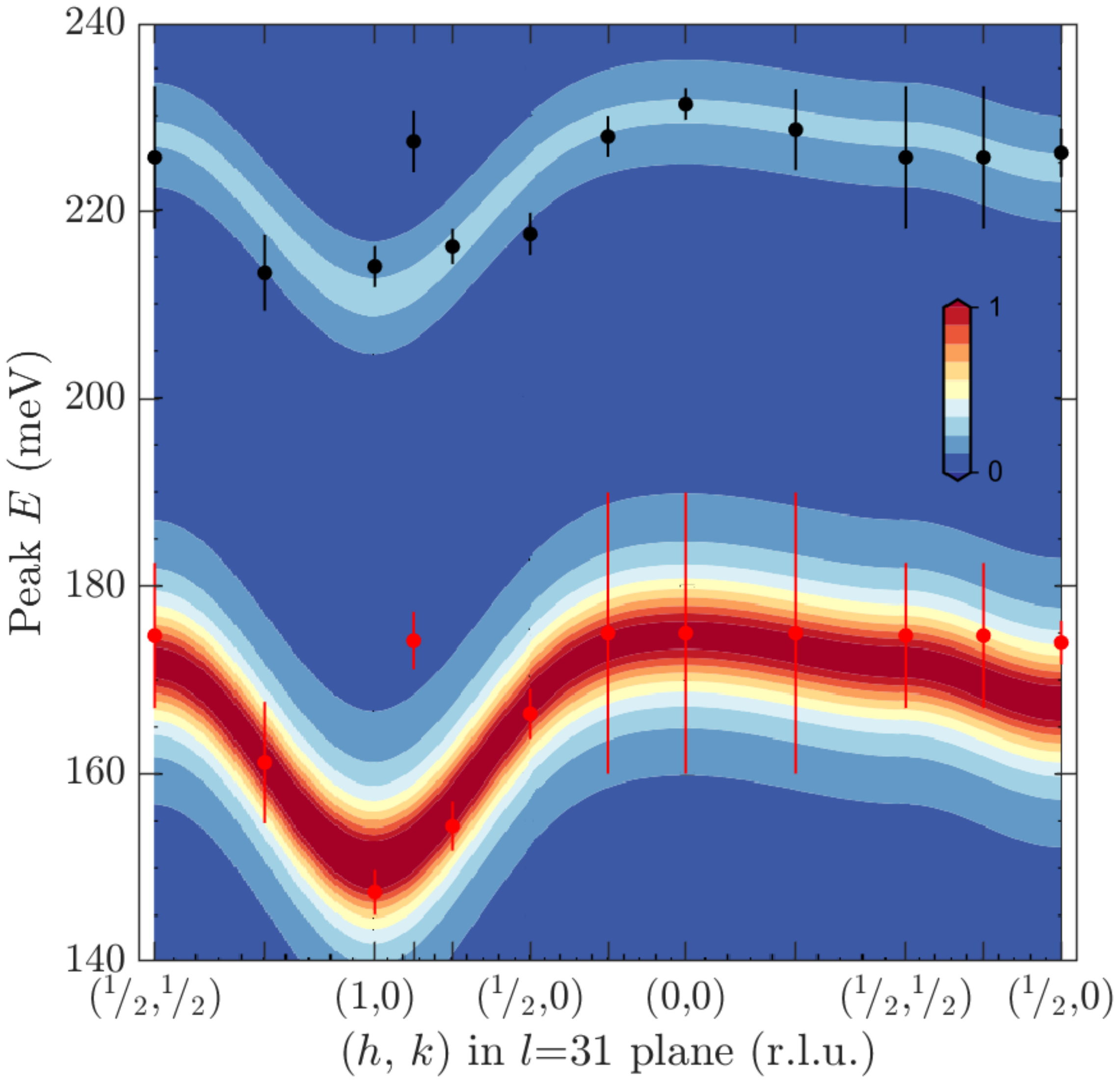}
\caption{Spin-orbit exciton model (solid colors and lines) in comparison to the fitted data (points). Energy transfers of transverse (lower energy) and longitudinal (higher energy) modes overplotted on $S(\mathbf{Q})$. Note that the RIXS intensity is not equivalent to $S(\mathbf{Q})$, and intensities have not been scaled to account for the scattering geometry. Error bars only account for peak energy uncertainty, and do not incorporate fixed widths ${\approx}60$ meV.} %(b) Effects of altering $\alpha'\lambda$ and $\delta$ for the $[-h,h]$ cut of the Brillouin zone, from ($1/2,1/2$) to ($1,0$).}
\label{fig:fig3}
\end{figure}

The energies of the peaks A, B can be determined under fixed $l$ and plotted across the 2D zone, and the resulting dispersion and level spacing is shown in Figure \ref{fig:fig3}.  This quasi-2D dispersion can then be phenomenologically fit via a spin-orbit exciton model assuming an idealized $S$=1 ($L$=1, $J$=0) ground state and a single-ion Hamiltonian $\hat{\mathcal{H}}_{\rm{S.I.}}$. This model for \sr~employs the same formalism that was established~\cite{Sarte20:102} for the $d^{4}$ multiorbital Mott insulator Ca$_{2}$RuO$_{4}$. This Hamiltonian is parameterized by: $\alpha'\lambda$, ${\rm{H}}_{MF}$, and $\delta$, corresponding to the individual contributions from spin-orbit coupling (with the prefactor $\alpha'{=}1/2$), an internal mean molecular field, and a uniaxial (either tetragonal or trigonal) distortion of the local octahedral coordination environment,  respectively. The resulting eigenstates of $\hat{\mathcal{H}}_{\rm{S.I.}}$ are then coupled by the Fourier transform of the exchange interaction $J(\mathbf{Q})$, where both an isotropic nearest neighbor $J_{1}$ and next nearest neighbor exchange $J_{2}$ are considered \cite{ESI}.

\newcolumntype{.}{D{.}{.}{4}}
%\newcolumntype{,}{D{,}{,}{4}}
\newcommand*{\thead}[1]{\multicolumn{1}{c}{\mdseries #1}}
\begin{table}[htb!]
\caption{Refined parameter values of the spin-orbit exciton model for \sr. All values are reported in meV.} 
	\begin{ruledtabular}
		\begin{tabular}{l.r.}
			\thead{Parameter} & \thead{Initial Value} & \thead{Range} & \thead{Refined Value}\\ 
			 \hline \\[-1em]
			$\lambda$ & 400 & [300, 500] & 374 \\
			$H_{MF}$ & 0 & [$-$5, $\quad$5] & 0.2  \\
			$\delta$ & 80 & [40, 120] &  100\\
			$J_{1}$ & 2& [0, $\quad$5]& 2.1 \\
			$J_{2}$ & -1 & [$-$2, $\quad$2] & -0.8 \\ 
	\end{tabular}
	\end{ruledtabular}
	\label{tab:3}
\end{table}

The refined parameters from this model are shown in Table~\ref{tab:3}, and the model's results are overplotted with the experimental data points in Figure~\ref{fig:fig3}. Two distinct modes are parameterized with calculated dispersion relations in excellent agreement with the experimental data. The lower energy A mode corresponds to transverse fluctuations ($\alpha\beta = +-$ and $-+$) within the basal plane of the pseudo-tetragonal unit cell, whereas the B mode at higher energy transfers corresponds to longitudinal $zz$ fluctuations along the Ir$^{5+}$ moment's axis.

\begin{table*}[t]
    \caption{Ir$^{5+}$ $5d^4$ multiplet structure, MRCI (without and with SOC), and RIXS relative energies, all in meV, for an embedded [IrO$_6$]$^{7-}$ octahedron using experimental crystallographic data. Notations reference cubic symmetry, even though the actual symmetry is much lower; for this reason, $T$ and $E$ crystal field states (2\textsuperscript{nd} column) are split up.}
    \begin{ruledtabular}
    \begin{tabular}{lllrl}
    Ir$^{5+}$ $t^4_{2g}$ terms & MRCI  & MRCI${+}$SOC   & RIXS            & Assignment \\
    \hline \\[-1em]
    $^3T_{1g}$           & 0     & 0        ($J\!\approx\!0$)  & 0               & Elastic ($S_z{=}$0)   \\
                         & 106   & 290, 370 ($J\!\approx\!1$)  & 170, 210        & A, B    ($S_z{=}\pm$1)\\
                         & 324   & 500      ($J\!\approx\!1$)  & $\approx$550    & C \\
                         &       & 720, 740, 760, 800, 870 ($J\!\approx\!2$)  & $\approx$750    & C \\
    $^1T_{2g}$, $^1E_g$  & 876, 896, 1013, 1235, 1341 ~~
                         & 1570, 1610, 1680, 1846, 1946 ($J\!\approx\!2$)     & $>$1000 & intra-$t_\text{2g}$\\
    $^1A_{1g}$           & 2396   & 3102  &               &\\
\end{tabular}
\end{ruledtabular}
\label{tab:qc}
\end{table*}

We now describe the quantitative results of the spin-orbit exciton model, with modeled uncertainties indicated in parentheses. The refined value of 374(10)~meV for the spin orbit coupling $\lambda$ is comparable to values reported for other $d^{4}$ iridates~\cite{Yuan2017,Kusch2018}. The presence of one, rather than two, transverse modes can be understood by the negligible molecular field. According to the model, $H\rm{_{MF}}$ has a refined value of 0.2(2)~meV, and this implies no splitting between the two plausible $\alpha\beta$= $+-$ and $-+$ transverse modes. The lack of a molecular field is also consistent with the absence of magnetic long-range order in this material. In such a case where the molecular field is absent, the gap between the longitudinal and transverse modes can be modeled via a uniaxial distortion of the coordination environment for a magnetic ion with unquenched orbital angular momentum. The large magnitude of $\delta$ with a refined value of 100(5)~meV yields a significant gap of $\delta/2$. The weak dispersion is captured by a weak antiferromagnetic $J_1{>}0$ and ferromagnetic $J_2{<}0$.  This model parameterizes these two lowest energy A and B excitations as transverse and longitudinal $S$=1 exciton branches, split by a strong tetragonal distortion, with negligible molecular field.

The spin-orbit exciton model described above assumes a simplified, uniform tetragonal distortion, which is not the case for \sr. As a complementary computational approach to the $5d$ electronic structure, quantum chemical calculations were carried out; see Table~\ref{tab:qc}. These \textit{ab initio} results indicate strong low-symmetry fields, comparable to the magnitude of SOC. The computed spin-orbit ground-state wavefunction features dominant in-plane orbital occupation and we confirm that the first two excited states essentially correspond to $S_\text{z}{=}{\pm}1$, although their relative energy is overestimated in the quantum chemical treatment. Another distinct level is found near 500\:meV. Without SOC, the lowest $S{=}0$ ($t^4_\text{2g}$) states are computed at $\approx$1\:eV (2\textsuperscript{nd} column in Table\:\ref{tab:qc}). Spin-orbit interactions push those levels to higher energies (3\textsuperscript{rd} column in Table\:\ref{tab:qc}), but these effects might be overestimated in the computations since higher-lying $t^3_\text{2g}e^1_\text{g}$ terms and O-to-Ir charge-transfer states are not included in the CAS treatment. As an approximation, additionally including the unoccupied $e_\text{g}$ atomic orbitals in the CAS treatment did not result in a closer agreement \cite{ESI}.

Despite overestimating the energy range of the $S_\text{z}{=}{\pm}1$ modes, the \textit{ab initio} data describes the RIXS spectrum qualitatively quite well. Most importantly, the energy splitting of the A and B modes modeled in the phenomenological spin-orbit exciton model is captured and the higher energy electronic transitions are identified. The A and B features in the RIXS spectrum can then be analyzed in terms of intramultiplet spin-orbit exciton modes within the group of $S{=}1$ ($J{\approx}0$ and $J{\approx}1$) states.

The full width at half maximum (FWHM) of the A and B exciton modes are reasonably narrow (60\:meV), close to the experimental resolution (35\:meV), allowing for a reliable extraction of the effective bandwidth: 15$-$35\:meV. Quantum chemical calculations of the [Ir$_2$O$_{11}$]$^{12-}$ embedded cluster model reveal that each of these modes is split by about 10\:meV. This is consistent with the dispersion arising from the exchange coupling. The C feature has FWHM values near 100 meV, and a bandwidth of about 80 meV, reminiscent of the broad spin-orbit excitons identified for Sr$_2$IrO$_4$ \cite{Kim2012Excitation} and Sr$_3$Ir$_2$O$_7$ \cite{Kim2014Excitonic}.

Having established that Sr$_3$Ir$_2$O$_7$F$_2$ provides a platform for the direct resolution of spin-orbit exciton modes in a $d^4$ iridate bilayer, it allows analysis of the anomalous excitation spectra previously reported in the Sr$_3$(Ir$_{1-x}$Ru$_x$)$_2$O$_7$ system.  Prior studies reported an unusual hardening of the magnetic excitation gap upon hole-doping and the persistence of robust antiferromagnetic spin fluctuations up to dopings as high as $x{=}0.77$ in Ruddlesden-Popper series strontium iridates \cite{Cao2017, Schmehr2019}.  The excitation bandwidth correspondingly narrows and localizes to ${\approx}200$ meV, the regime where the S$_z{=}{\pm}1$ modes are observed in Sr$_3$Ir$_2$O$_7$F$_2$.

Prior models leveraged local exchange disorder or coupling to the particle-hole continuum to explain the anomalous excitation spectra of Sr$_3$(Ir$_{1-x}$Ru$_x$)$_2$O$_7$; however our resolution of the purely excitonic spectrum of Sr$_3$Ir$_2$O$_7$F$_2$ should map to the $d^4$ endpoint of Sr$_3$(Ir$_{1-x}$Ru$_x$)$_2$O$_7$.  Our findings suggest that local $d^4$ iridium sites in the electronically inhomogeneous iridate-ruthenate alloy~\cite{dhital2014carrier} likely drive the apparent gap hardening.  The lifetimes of low energy excitations in Sr$_3$(Ir$_{1-x}$Ru$_x$)$_2$O$_7$ are exceptionally broad, and as the magnetic moment collapses with hole-doping upon entering the metallic regime, the local spin-orbit excitons endemic to $d^4$ sites dominate the RIXS spectrum.   

Sr$_3$Ir$_2$O$_7$ itself has recently been proposed to be an excitonic insulator with an antiferromagnetic ground state \cite{mazzone2022antiferromagnetic}, and Sr$_3$Ir$_2$O$_7$F$_2$ provides an important nonmagnetic $J{=}0$ comparator for understanding how the local electronic structure and interactions evolve as holes are introduced into the compound. Our RIXS data combined with quantum chemistry calculations and a phenomenological spin-orbit exciton model allow us to identify the effective spin-orbit coupling, tetragonal distortion, and exchange coupling energies in a unique, $d^4$ Ruddelsden-Popper bilayer iridate.

%%%%%%%%%%%%%%%%%%%%%%%%%%%%%%%%%%%%%%%%%%%%%%%%%%%%%

\begin{acknowledgments}
The authors acknowledge useful discussions with 
A.~Revelli, T.E.~Mates, C.~Stock, H.~Lane, B.R.~Ortiz, K.J.~Camacho, U.K.~R{\"o}{\ss}ler, S.~Nishimoto, 
and C.~Schwenk. This work was supported by NSF award DMR-1729489 (S.D.W. and Z.P.). P.M.S. acknowledges additional financial support from the University of California, Santa Barbara, through the Elings Fellowship. T.P. and L.H. acknowledge 
U.~Nitzsche for technical assistance and financial support from the German Research Foundation (Deutsche Forschungsgemeinschaft, DFG), Project Nos. 437124857 and 468093414. The research made use of the shared facilities of the NSF Materials Research Science and Engineering Center at UC Santa Barbara, Grant No. DMR-1720256. We also acknowledge the use of the Nanostructures Cleanroom Facility within the California NanoSystems Institute, supported by the University of California, Santa Barbara and the University of California, Office of the President. Use of the Advanced Photon Source at Argonne National Laboratory was supported by the U. S. Department of Energy, Office of Science, Office of Basic Energy Sciences, under Contract No. DE-AC02-06CH11357.   
\end{acknowledgments}

%%%%%%%%%%%%%%%%%%%%%%%%%%%%%%%%%%%%%%%%%%%%%%%%%%%%%

\bibliography{sr3ir2o7f2_rixs_bib}

%%%%%%%%%%%%%%%%%%%%%%%%%%%%%%%%%%%%%%%%%%%%%%%%%%%%%
\end{document}